 \documentstyle[fullpage,12pt]{article}
\begin{document}

 \baselineskip 13pt
 \parskip 4pt

\def\cl{\centerline}
\def\bk{\hfill\break}
\def\no{\noindent}
\def\etal{{\it et al.\ }}
\def\rms{{\it rms\ }}
\def\cf{{\it cf.\ }}
\def\eg{{\it e.g.}}
\def\ie{{\it i.e.}}
\def\ltsima{$\; \buildrel < \over \sim \;$}
\def\lsim{\lower.5ex\hbox{\ltsima}}
\def\gtsima{$\; \buildrel > \over \sim \;$}
\def\gsim{\lower.5ex\hbox{\gtsima}}
\def\kms{\ {\rm km\,s^{-1}}}
\def\gcc{\,{\rm g\,cm^{-3}}}
\def\pcc{\,{\rm cm^{-3}}}
\def\hmpc{\,{ h^{-1}{\rm Mpc} }}
\def\mpc{\,{\rm Mpc}}
\def\kmsmpc{\,{\rm km\,s^{-1}Mpc^{-1}}}
\def\ln{{\rm ln}}
\def\log{{\rm log}}
\def\cos{{\rm cos}}
\def\sin{{\rm sin}}
\def\tan{{\rm tan}}
\def\dd{{d}}
\def\pa{\partial}
\def\rarrow{\rightarrow} \def\larrow{\leftarrow}
\def\la{\langle} \def\ra{\rangle}
\def\solar{\ifmmode_{\mathord\odot}\;\else$_{\mathord\odot}\;$\fi}
\def\msun{\rm M_{\solar}\!}
\def\lsun{\rm L_{\solar}\!}
%
\def\pmb#1{\setbox0=\hbox{#1}%
 \kern-.025em\copy0\kern-\wd0
 \kern.05em\copy0\kern-\wd0
 \kern-.025em\raise.0433em\box0}
\def\vv{\pmb{$v$}}
\def\vx{\pmb{$x$}}
\def\vr{\pmb{$r$}}
\def\vnabla{\pmb{$\nabla$}}
\def\divv{\vnabla\!\cdot\!\vv}
%
\def\jo{\it}
\def\apj#1{{\jo Ap.J.}~{\bf #1}}
\def\apjl#1{{\jo Ap.J. Lett.}~{\bf #1}}
\def\apjs#1{{\jo Ap.J. Supp.}~{\bf #1}}
\def\mn#1{{\jo MNRAS}~{\bf #1}}
\def\aa#1{{\jo Astron. Astrophys.}~{\bf #1}}
\def\aj#1{{\jo Astron. J.}~{\bf #1}}
\def\nat#1{{\jo Nature}~{\bf #1}}
\def\araa#1{{\jo Annu. Rev. Astron. Astrophys.}~{\bf #1}}
\def\pasp#1{{\jo Proc. Ast. Soc. Pac.}~{\bf #1}}
\def\rmp#1{{\jo Rev. Mod. Phys.}~{\bf #1}}
\def\prl#1{{\jo Phys. Rev. Lett.}~{\bf #1}}
\def\pl#1{{\jo Phys. Lett.}~{\bf #1}}
\def\plb#1{{\jo Phys. Lett. B}~{\bf #1}}
\def\pr#1{{\jo Phys. Rep.}~{\bf #1}}
\def\prd#1{{\jo Phys. Rev. D}~{\bf #1}}
\def\prb#1{{\jo Phys. Rev. B}~{\bf #1}}
\def\ca#1{{\jo Comments Astrophys.}~{\bf #1}}
\def\baas#1{{\jo Bull. Am. Astron. Soc.}~{\bf #1}}
\def\sc#1{{\jo Science}~{\bf #1}}
\def\omm{\Omega_m}
\def\oml{\Omega_\Lambda}
\def\omk{\Omega_k}
\def\omb{\Omega_b}
\def\omn{\Omega_\nu}
\def\omt{\Omega_{tot}}
\def\oma{\Omega_a}
\def\ho{H_0}
\def\to{t_0}

\def\dns{$D_n-\sigma$}
\def\h65{h_{65}}
\def\bi{b_{\scriptscriptstyle I}}
\def\betai{\beta_{\scriptscriptstyle I}}
\def\bo{b_{\scriptscriptstyle O}}
\def\betao{\beta_{\scriptscriptstyle O}}

 \title{{\bf Measuring Omega}\footnote{To appear in  
 Critical Dialogues in Cosmology (Princeton 250th Anniversary), 
 ed. N. Turok (World Scientific)}}

 \author{{\bf Avishai Dekel$^1$, David Burstein$^2$, Simon D.M. White$^3$}\\ 
 \\
$^1$The Hebrew University of Jerusalem\\
$^2$Arizona State University, Tempe\\
$^3$Max-Planck-Institut fuer Astrophysik, Garching}

\date{}

\maketitle

\section 
{Introduction}

We were asked to debate the value of $\Omega$, the fundamental energy density
parameter of cosmology, and in particular its mass component, $\omm$. 
Is the universe flat and marginally bound with 
$\omm=1$ in accordance with the simplest
cosmological model?  Is $\omm$ clearly smaller than unity
as seems to be indicated by several observations?
Unfortunately, we cannot provide a clear answer at this point
because there is conflicting evidence.
Entertaining the audience with our biased views
on the subject might not be very constructive.
Instead, it may be more interesting to
lay out the various {\it methods} used to measure $\omm$,
mention new developments and current estimates,
and focus on the promising prospects versus the associated difficulties.
In the critical discussion that follows we try to shed light on the nature
of the uncertainties that may be responsible for the current span of 
estimates for $\omm$.

We divide the methods into the following four classes:

\leftskip=1.0 true cm

\leftskip=1.0 true cm
\noindent
{\hskip -0.5 true cm}
$\bullet\ ${\it Global measures}. Based on properties of space-time
that constrain combinations of $\omm$ and the
other cosmological parameters ($\Lambda$, $H_0$, $t_0$).

\leftskip=1.0 true cm
\noindent
{\hskip -0.5 true cm}
$\bullet\ ${\it Virialized Systems}. Methods based on nonlinear dynamics 
within galaxies and clusters on comoving scales $1-10\hmpc$.

\leftskip=1.0 true cm
\noindent
{\hskip -0.5 true cm}
$\bullet\ ${\it Large-scale structure}. Measurements based on 
mildly-nonlinear gravitational dynamics of fluctuations on scales 
$10-100\hmpc$ of superclusters and voids, in particular {\it cosmic flows}.

\leftskip=1.0 true cm
\noindent
{\hskip -0.50 true cm}
$\bullet\ ${\it Growth rate of fluctuations}. Comparisons of present day
structure with fluctuations at the last scattering of the cosmic microwave 
background (CMB) or with high redshift objects of the young universe.

\leftskip=0.0 true cm

The methods and current estimates are discussed below 
and summarized in Figure 1 and Table 1.
The estimates based on virialized objects typically yield low values of
$\omm \sim 0.2-0.3$. The global measures, large-scale structure and cosmic
flows typically indicate higher values of $\omm \sim 0.4-1$.

\section 
{Global Measures}

We adopt as our basic working hypothesis 
the standard cosmological model of Friedmann Robertson Walker (FRW),
where we assume homogeneity and isotropy and describe 
gravity by general relativity. We limit the discussion to the matter-dominated 
era in the ``dust'' approximation.

The Friedmann equation that governs the universal expansion
can be written in terms of the different contributions to
the energy density (\eg, \cite{carroll92}):
\begin{displaymath}
\omm +\oml +\omk = 1 \ ,
\end{displaymath}
\begin{equation}
\omm\equiv {\rho_m \over (3H^2/8\pi G)}, \quad
\oml\equiv {\Lambda c^2 \over 3H^2}, \quad
\omk\equiv {-k c^2 \over a^2 H^2} \ .
\label{eq:omegas}
\end{equation}
Here, $a(t)$ is the expansion factor of the universe, 
$H(t)\equiv \dot{a}/a$ is the Hubble constant,
$\rho_m(t)$ is mean the mass density,
$\Lambda$ is the cosmological constant,
and $k$ is the curvature parameter.
Hereafter, the above symbols for the cosmological parameters
refer to their values at the present time, $t_0$.

We denote $\omt\equiv \omm+\oml$, which by Eq. \ref{eq:omegas} equals 
$1-\omk$;  its value relative to unity determines whether the universe 
is open ($k\!=\!-1$), flat ($k\!=\!0$), or closed ($k\!=\!+1$).
Another quantity of interest is the deceleration parameter, 
$q_0\equiv -a \ddot{a} / {\dot a}^2$, which by Eq. \ref{eq:omegas} 
is related to the other parameters via
$q_0 = \omm/2 - \oml$.

The FRW model also predicts a relation between the dimensionless product
$\ho\to$ and the parameters $\omm$ and $\oml$. 
For $\oml=0$, this product ranges between
1 and 2/3 for $\omm$ in the range 0 to 1 respectively, and it is
computable for any values of $\omm$ and $\oml$ (\S \ref{sec-age}).

The global measures commonly involve combinations of the cosmological
parameters. Constraints in the $\omm-\oml$ plane
are displayed in Figure \ref{fig:omlam}.

\subsection 
{Occam's Razor}

The above working hypotheses, and the order by which more specific 
models should be considered against observations, are guided by the 
principle of Occam's Razor, \ie, by simplicity and robustness to initial
conditions. The caveat is that different researchers might disagree 
on the evaluation of ``simplicity''.

It is commonly assumed that the
simplest model is the Einstein-deSitter
model, $\omm=1$ and $\oml=0$. One property that makes it robust is the
fact that $\omm$ remains constant
at all times with no need for fine tuning at the initial conditions
(the ``coincidence'' argument \cite{bondi60}).
 
The most natural extension according to the
generic model of inflation is a flat universe, $\omt=1$, where
$\omm$ can be smaller than unity but only at the expense of a 
nonzero cosmological constant.
 
These simple models could serve as useful references, and even guide the
interpretation of the results, but they should not bias the measurements.

\subsection 
{Classical Tests of Geometry}

The parameter-dependent large-scale geometry of space-time
is reflected in the volume-redshift relation.
There are two classical versions of the tests that utilize this dependence:
magnitude versus redshift (or ``Hubble diagram")
and number density versus redshift.
The luminosity distance to a redshift $z$, 
which enters the Hubble diagram test,
depends on $\omm$ and $\oml$ via the integral (\eg, \cite{carroll92})
\begin{displaymath}
d_l(z) = {c (1+z)\over H_0 \vert\omk\vert^{1/2} }\
S_k \left[\,\vert\omk\vert^{1/2} \int_0^z F(\omm,\oml,z')\, dz' \,\right] \ , 
\end{displaymath}
\begin{equation} 
F(\omm,\oml,z) \equiv [(1+z)^2 (1+\omm z) -z(2+z)\oml ]^{-1/2} \ ,
\label{eq:dl}
\end{equation}
where $S_0(x)\equiv x$, $S_{+1}\equiv \sin$ and $S_{-1}\equiv \sinh$.
At $z \sim 0.4$, $d_l$ happens to be (to a good approximation)
a function of the combination $\omm-\oml$ (not $q_0$) \cite{perlmutter96}. 
The angular diameter distance, which enters the tests based on number density,
is simply $d_a = d_l / (1+z)^2$.

{\it New Developments:}
Accumulating data of supernovae type Ia (SNIa) 
out to $z\sim 0.4$ and beyond look promising for a
Hubble-diagram test \cite{perlmutter96}.
The preliminary success of the method may indicate that it will be able
to separate the dependences on $\omm$ and $\oml$ within a few years,
once several supernovae are measured at $z\sim 1$ \cite{goobar95}.
$\bullet$
Measurements of the galaxy number count $N(m,z)$ seem to be in
reach for surveys at high redshift \cite{lilly96}.

{\it Pro:}
The main advantage of such tests is that they are
direct measures of global geometry and thus independent
of assumptions regarding the mass type and distribution, the
statistical nature of the fluctuations, the growth by gravitational
instability (GI) and galaxy biasing.
$\bullet$
The galaxy-type ``standard candles" that were used over the years
clearly suffer from severe evolution complications.
Supernovae type Ia are the popular current candidate for a standard
candle, based on the assumption that
stellar processes are not likely to vary much in time.
$\bullet$
Systematic searches for supernovae are in progress.

{\it Con:}
The key question is whether SNIa are indeed a standard
candle.  Some caution is in place as long as we lack
a complete theory for supernovae.
If they are exploding white dwarfs, perhaps the generic SNIa
at $z\sim 1$ comes from a higher mass white dwarf than one does today?
$\bullet$
Luminosity density distributions also have to assume 
how galaxies evolve.  If galaxies are formed in a series of hierarchical
mergers that continues at low levels today, there will be more galaxies
in the past than now, requiring an accurate theory of galaxy merging to
deduce an accurate estimate of density evolution linked to cosmology. 

{\it Current Results:}
The first 7 supernovae analyzed by Perlmutter \etal \cite{perlmutter96}
at $z\sim 0.4$ yield $-0.3 < \omm-\oml < 2.5$ as the 90\% two-parameter
likelihood contour (Fig. \ref{fig:omlam}).
For a {\it flat} universe they find for each parameter
$\omm=0.94^{+0.34}_{-0.28}$, and
$\oml < 0.51$ (or $\omm > 0.49$) at 95\% confidence.
Improved results are expected soon from tens of supernovae.
$\bullet$
So far, the galaxy number counts from the Hubble Space Telescope (HST) 
and the 10-meter Keck telescope still yield conflicting results
(see a summary in \cite{peebles97}).

\subsection 
{Number Count of Quasar Lensing}

This is a promising new version of the classical number density test.
Strong sensitivity to $\oml$ arises when $\oml$ is positive and comparable 
to $\omm$. In this case, the universe should have
gone through a phase of slower expansion in the recent cosmological
past, which should be observed as an accumulation of objects at a specific
redshift of order unity.
In particular, it should be reflected in the observed
rate of lensing of high-redshift quasars by foreground galaxies 
\cite{fukugita90}.
The probability of lensing of a source at redshift $z_s$
by a population of isothermal spheres of constant comoving density
as a function of the cosmological parameters is \cite{carroll92}:
\begin{equation}
P_{lens} \propto \int_0^{z_s}
  (1+z)^2 F(\omm,\oml,z)\, [ d_a(0,z) d_a(z,z_s) / d_a(0,z_s)] ^2\, dz \ ,
\label{eq:lens}
\end{equation}
where $d_a(z_1,z_2)$ is the angular diameter distance from $z_1$ to $z_2$.
The contours of constant lensing probability in the $\omm-\oml$ plane
for $z_s \sim 2$ \cite{carroll92} 
happen to almost coincide with the lines $\omm-\oml=const.$ 
The limits from lensing are thus similar in nature to the limits
from SNIa.

{\it Pro:}
This test shares all the advantages of direct geometrical measures, \eg, being
independent of dark matter, fluctuations, GI and galaxy biasing.
The high redshifts involved bring about a unique sensitivity to $\oml$,
compared to the negligible effect that $\oml$ has on the structure observed
at $z\ll 1$.

{\it Con:}
The constraint is weakened if the
lensed images are obscured by dust in the early-type galaxies 
that are responsible for the lensing, especially 
if E galaxies at $z\sim1$ have nuch more dust than low-redshift ones 
\cite{malhotra97}. There is no clear evidence for a strong effect.
$\bullet$ 
A similar uncertainty arises 
if these galaxies had rapid evolution between 
$z\sim 1$ and the present; current evidence suggests a weak evolution.
$\bullet$
The method was criticized for it's sensitivity 
to the velocity dispersion assumed for the typical lenses and thus to the 
galaxy luminosity function \cite{chiba96}, but the requirement that the 
distribution of lenses should simultaneously produce the observed 
distribution of image separations largely
invalidates this criticism \cite{kochanek96}.

{\it Current Results:} From 
the failure to detect the accumulation of lenses, the current limit for a 
{\it flat} model is $\oml < 0.66$ (or $\omm > 0.36$)
at 95\% confidence \cite{kochanek96} (Fig. \ref{fig:omlam}).
If $\oml=0$, this test provides only a weaker bound, $\omm>0.2$
at 90\% confidence. Several new lenses are found each year, promising slow
but continuous improvement.

\subsection 
{Microwave Background Acoustic Peaks}

This is the most promising test. In less than a decade it is expected
to provide the most stringent constraints on the cosmological parameters.
The test uses the effect of the background cosmology on the geodesics 
of photons.
Current ground-based and balloon-born experiments already provide 
preliminary constraints on the {\it location}
of the first acoustic peak on sub-degree scales in the angular power
spectrum of CMB temperature fluctuations,  $l(l+1)C_l$. The dependence
of the peak location on the
cosmological parameters enters via the combined effect of
(a) the physical scale of the ``sound horizon" that is
proportional to the cosmological horizon at recombination,
and (b) the geometry of space-time via the angular-diameter
distance. In the vicinity of a flat model,
the first peak is predicted at approximately the multipole (\eg, \cite{silk}
\cite{white-scott96})

\begin{equation}
l_{peak} \simeq 220 (\omm+\oml)^{-1/2} .
\label{lpeak}
\end{equation}

{\it New Developments:}
The next generation of post-COBE CMB satellites (MAP to be launched by
NASA in 2001, and in particular COBRAS/SAMBA scheduled by ESA for 2004) 
are planned to
obtain a precision at $\sim 10$ arc-minute resolution
that will either rule out the current framework of GI
for structure formation or
will measure the cosmological parameters to high precision.
Detailed evaluation of COBRAS/SAMBA shows that
nominal performance and expected foreground subtraction noise will
allow parameter estimation with the following accuracy:
$H_0 \pm 1\%$, $\omt \pm 0.005$, $\oml \pm 0.02$, $\omb \pm 2\%$.

{\it Pro:}
The precision hoped for is much better than attainable with any other known
method. If the observations fit the model, the precision is such that
the model will be confirmed beyond reasonable doubt.
$\bullet$
The constraints on $\omt$ come mostly from geometrical effects.
The interpretation is based on well understood physics of sound waves
in the linear regime, and on the assumption of absence of any relevant
preferred scale (in the megaparsecs to gigaparsec range)
in the physics which generated the
initial structure. The latter assumption can be checked directly by the
observations themselves.

{\it Con:}
The measurements might be messed up by unexpected foreground
contamination (\eg, by diffuse matter in galaxy groups).
$\bullet$
The detailed measurements need to wait 5 to 10 years.
$\bullet$
The assumption of no preferred scale in the initial fluctuations
may be wrong.
$\bullet$
If the observations do not fit the model, the whole
paradigm behind current structure formation modeling will be excluded,
and then no parameter estimates will be possible.
However, this seems unlikely given the recent measurements from the
ground and from balloons, which have already confirmed the
existence of the first acoustic peak.

\begin{figure}[t!]
 \centerline{\epsfxsize=5.0 in \epsfbox{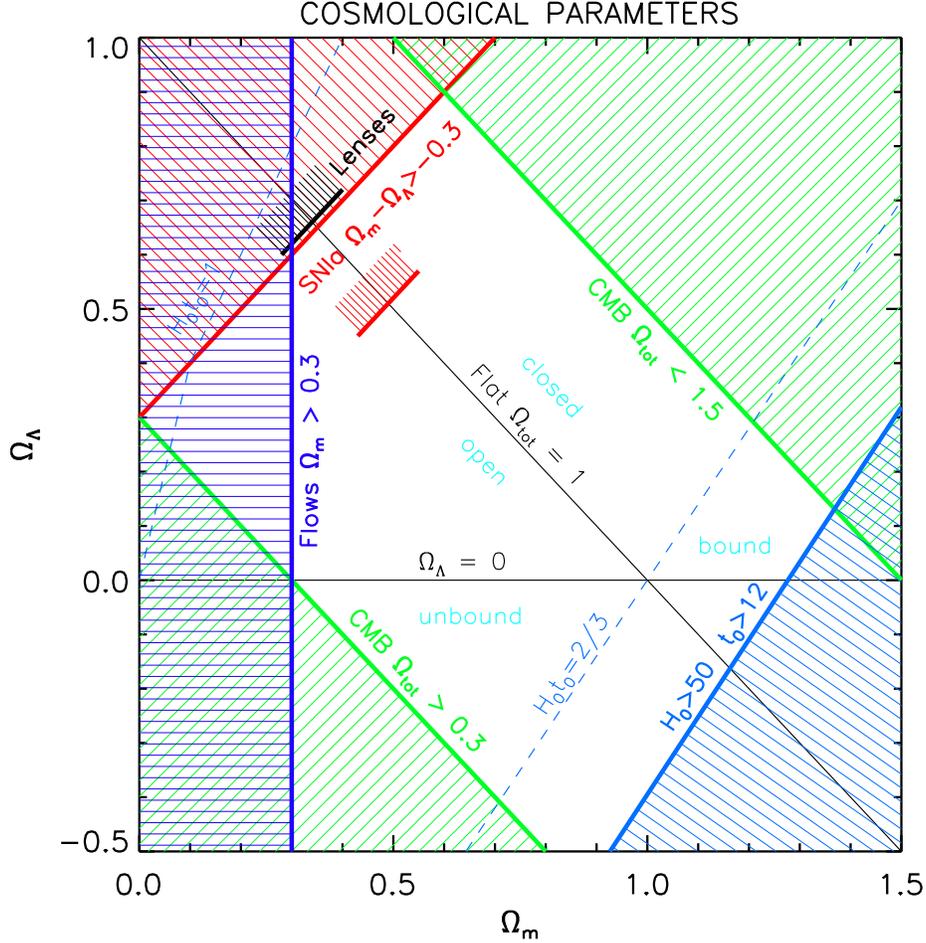}}
\caption{{\small
Current limits ($\sim 2\sigma$)
on the cosmological parameters $\omm$ and $\oml$ from
global measures: luminosity distance of SNIa, lens count,
the location of the CMB peak, and the age versus Hubble constant.
The short marks are the one-parameter 95\% limits from SNIa and lenses 
for a flat universe. 
Also shown (vertical line) is the 95\% lower bound on $\omm$ from cosmic flows.
The most likely value of $\omm$ lies in the range 0.5 to 1.
The Einstein-deSitter model is permitted.  
An open model with $\omm\simeq0.2$ and $\oml=0$, 
or a flat model with $\omm\simeq 0.3$ and $\oml\simeq0.7$, 
are ruled out.
}}
\label{fig:omlam}
\end{figure}

{\it Current Results:}
Balloon and ground-based results have already confirmed the
existence of the first acoustic peak, and have constrained it's location
to the vicinity $l\sim 200$.
The results of COBE's DMR ($l\sim 10$) provide an upper bound of
$\omm+\oml < 1.5$ at the 95\% confidence level 
for a scale-invariant initial spectrum
(and the constraint becomes tighter for any ``redder" spectrum, $n<1$)
\cite{white-scott96}.
Several balloon experiments ($l\sim 50-200$) 
strengthen this upper bound \cite{cmb_degree}.
The CAT experiment ($l\sim 350-700$)  
yields a preliminary lower bound of $\omm+\oml > 0.3$ \cite{cat} 
(Fig. \ref{fig:omlam}).

\input epsf

\subsection 
{The Age of the Universe}
\label{sec-age}

Measured independent lower bounds on the Hubble constant and on the age
of the oldest globular clusters provide a lower bound on $\ho\to$ 
($=1.05 ht$, where $H_0 \equiv 100 h \kmsmpc$ and $t_0\equiv 10 t {\rm Gyr}$),
and thus an interesting constraint in the $\omm-\oml$ plane.
The exact expressions are computable in the various regions of
parameter space.
For example, for $\oml=0$, the relation is
(\eg,  \cite{padma93}, Eq. 2.79)
\begin{equation}
\ho\to= {1\over2}\, {\omm \over \vert 1-\omm \vert ^{3/2}}\, k\,
      \left[ C_k^{-1} \left( {2\over \omm}-1 \right) 
             -{2 \vert 1-\omm \vert ^{1/2} \over\omm} \right] \ ,
\label{eq:ht_nolambda}
\end{equation}
where $C_{+1}^{-1} \equiv \cos^{-1}$ and $C_{-1}^{-1} \equiv \cosh^{-1}$.
A very useful approximation in the presence of a cosmological constant,
that is an exact solution for a flat universe, is \cite{carroll92}
\begin{equation}
\ho\to = {2\over3} \, {1\over \vert 1-\oma \vert ^{1/2}} \,
     S_a^{-1} \left( {\vert 1-\oma \vert ^{1/2} \over \oma^{1/2}} \right),
\quad \oma \equiv 0.7 \omm - 0.3 \oml +0.3 \ ,
\label{eq:ht_lambda}
\end{equation}
where $S_{\oma \leq 1}^{-1} \equiv \sinh^{-1}$   
and $S_{\oma > 1}^{-1} \equiv \sin^{-1}$.
A useful crude approximation near $\ho\to \sim 2/3$ is
\begin{equation}
\omm -0.7\oml \simeq 5.8 (1- 1.3 ht) \ .
\label{eq:ht}
\end{equation}

{\it New Developments:}
Progress is being made in the HST key project to measure $\ho$ based on
Cepheids in Virgo and Fornax.

{\it Pro:}
The method does not depend on fluctuations, GI, biasing, etc.
$\bullet$
The current error of $\sim 20\%$ in the Hubble constant
will hopefully be reduced soon to the level of 10\% percent.
$\bullet$
This method is likely to provide the most stringent upper bound on
$\omm$ and lower bound on $\oml$.

{\it Con:}
The errors in the determination of the age of the universe based on
globular clusters are uncertain. The major source of error are the
distances to Pop--II stars in the globular clusters, and
complex stellar evolution issues.

{\it Current Results:}
The most likely estimates are of $h\simeq 0.6-0.7$ \cite{hubble}
and $t\simeq 1.5$ \cite{age},
corresponding to $ht\simeq1$. They seem to favor a possible
deviation from the Einstein deSitter model towards low $\omm$
or high $\oml$ or both.
However, one only needs to appeal to the current $\sim1\sigma$ lower bounds
(say $h \simeq 0.53$ and $t \simeq 1.2$)
in order to accommodate the Einstein deSitter model.

\medskip
Figure \ref{fig:omlam} displays in the $\omm$-$\oml$ plane the $\sim$ two-sigma 
constraints from the global measures discussed above.
Superposed is the main constraint from cosmic flows (\S \ref{sec-flows} below).
The joint permitted range for $\omm$ is thus roughly 0.4 to 1.1.  
Low $\omm$ models of $\omm \leq 0.3$ are significantly ruled out.

\section 
{Virialized Objects: Galaxies and Clusters}

These methods measure the mass associated with galaxies and
clusters on scales of order one to a few $\hmpc$.
The structure is assumed to have grown from small initial fluctuations
via GI, and the methods involve nonlinear dynamics of virialized systems.
The cosmological constant is irrelevant; the structure at
low redshift is insensitive to $\oml$.

Several additional physical quantities are now involved, such as
the baryonic contribution to the universal mass density, $\omb$.
The complex relation between galaxies and mass, termed ``biasing",
enters as a crucial unknown on these galactic scales,
and issues of gas dynamics are important. These complications introduce 
severe uncertainties into the estimates of $\omm$.

\subsection 
{Mass to Light Ratio}

The mass-to-light ratio $M/L$ is measured from virialized
systems from galaxies up to rich clusters.
The asymptotic universal value of the luminosity density,
${\cal L}$, is estimated by integrating the galaxy luminosity function.
The mass-to-light ratio in clusters is assumed to be the same
as outside clusters. Then
\begin{equation}
\omm = 0.33 \left( {M/L\over 300 h \msun/\lsun} \right)
       \left( {{\cal L} \over 3\times 10^8 h \lsun \mpc^{-3} } \right) \ .
\label{eq:moverl}
\end{equation}

{\it New Developments:}
Better masses for clusters and groups are now available
from virial analysis of galaxies, from X-ray gas
with resolved temperature structure, and from gravitational lensing.
$\bullet$
Masses for extended galaxy halos are also becoming available
from satellite galaxies and from lensing.

{\it Pro:}
This is a simple, straightforward method.
The result is independent of $\ho$.

{\it Con:}
The basic hypothesis is unjustified. It rests on the implicit assumption that
galaxies form in an unbiased way. Clusters and groups may have extended
dark halos that are not traced by the galaxies or by the X-ray gas, so $M/L$
may vary within the clusters as a function of radius.  This problem is
even worse within galactic halos.
$\bullet$
The total mass of clusters may be underestimated in the virial analysis based
on galaxy velocities because of possible velocity antibiasing, and because of
applying a spherical analysis to elongated systems.
$\bullet$
The mass based on X-ray measurements may be underestimated by the
assumption that the X-ray gas is in hydrostatic equilibrium.
The data allows significant freedom in the mass profile,
which translates to an uncertainty of factor two
within the Abell radius, and a larger error at larger radii \cite{balland}.
$\bullet$
The luminosity density may not be contributed by the same stars
that dominate the cluster light.
There are factors of 2-3 uncertainties in ${\cal L}$
stemming, for example, by differences in overall metallicity
and extinction by dust.

{\it Current Results:} From 
virial analysis of galaxy velocity dispersions in several clusters
stuck together, and a corresponding estimate of $\cal L$:
$\omm \simeq 0.25 \pm 0.05$ \cite{carlberg96}.

\subsection  
{Baryon Fraction and Nucleosynthesis}

The baryon fraction in clusters, $f_b$, is assumed to be equal to
the cosmic value. Combined with the universal baryon density $\omb$
as predicted from light element abundances through the theory of
big-bang nucleosynthesis (BBN) it yields the cosmic mass density 
\cite{white_fb93}:

\begin{equation}
\omm = (\omb/0.1) (f_b/0.1)^{-1} .
\label{eq:baryonfrac}
\end{equation}

{\it New Developments:}
The baryonic content is mostly in the form of X-ray
gas, and is estimated from ROSAT to large radii.
The reliability of the gas mass estimates is partly confirmed by
(a) direct spectral limits on gas inhomogeneity, and
(b) limits on the cluster magnetic fields from Faraday rotation of
background sources.
Independent cluster masses are obtained from gravitational lensing.
$\bullet$
The Deuterium abundance is being measured from quasar absorption systems
that are assumed to be composed of unprocessed primordial material.

{\it Pro:} This method avoids assumptions about galaxy biasing and stellar
populations; it depends on the
relatively safe assumption that cluster formation proceeds by
collapse from a well mixed medium, and the assertion that only little
segregation occurred between the gas and the dark matter.

{\it Con:}
The baryonic mass could be overestimated if the X-ray gas is
locally inhomogeneous, or if large tangled magnetic fields
provide a significant part of the pressure in cluster centers.
$\bullet$
The total mass of clusters may be underestimated as in the $M/L$ method.
$\bullet$
The baryon fraction seems to increases with cluster mass and
sometimes with radius within a cluster.
Plausible physical processes eject gas from the inner regions of lower mass
systems, making $f_b$ an underestimate of the global baryon
fraction, and leading to an overestimate of $\omm$.
The method assumes that the baryonic fraction in clusters is equal to
the universal value even though clusters only contain a few percent of the
galaxies in the universe.
$\bullet$
The method relies on the uncertain interpretation of the
light element abundances and on the theory of big-bang nucleosynthesis.

{\it Current Results:}
The baryonic fraction is estimated to be in the range 
$f_b=(0.03-0.08)h^{-3/2}$ by \cite{white_fb95}, but has also
been estimated to have a factor of 5 range among galaxy groups and
clusters of similar mass \cite{mushotzky96}.
The larger values typically apply to the most massive
and best observed clusters, but preliminary ASCA results indicate
lower values in several rich clusters. 
Low values are also indicated in groups \cite{mulchaey96}.
$\bullet$
The current BBN results suffer from uncertainty in the primordial
Deuterium abundance. The traditional estimates are of low $\omb$, \eg,
of $0.009 \leq \omb h^2 \leq 0.02$ based on all the light elements,
and $0.006 \leq \omb h^2 \leq 0.03$ from Deuterium only \cite{copi95} 
Recent, high-resolution spectra from the Keck telescope show a lower Deuterium 
abundance in two quasar absorption systems that seem to consist of 
unprocessed primordial material,
and correspondingly $\omb h^2 = 0.024_{-0.005}^{+0.006}$ \cite{tytler96}.

With $f_b$ in the middle of the range quoted above,
the estimate is either $\omm \simeq 0.3 \h65^{-1/2}$ (for low $\omb$) 
or $\omm \simeq 0.55 \h65^{-1/2}$ (for high $\omb$),
but in either case $\omm=1$ cannot be definitively excluded.

\subsection 
{Cosmic Virial Theorem}

In the cosmic virial theorem (CVT), clustering is assumed to be
statistically stable on scales $\lsim 1\hmpc$, such that the ensemble-average
contribution of a third mass particle to the relative acceleration of
a pair of galaxies is balanced by the relative motions of the ensemble of
pairs.  The kinetic energy is represented
by the pairwise relative velocity dispersion of galaxies [$\sigma_{12}(r)$],
and the potential energy involves a combination of integrals of the
2- and 3-point galaxy correlation functions \cite{peebles80}.
A very crude approximation to the actual relation is

\begin{equation}
\sigma_{12}^2(r) = \omm \xi(r) r^2 ,
\label{eq:cvt}
\end{equation}
where $\xi(r)$ is the two-point correlation function.
In some ways this is like
stacking many groups of galaxies to get a mean $M/L$,
so it can be regarded as a statistical version of the mass-to-light
ratio method.

On scales $1-10\hmpc$ the more useful statistics for dynamical mass estimates
is the mean (first moment) pairwise velocity in comparison with an integral
of the galaxy two-point correlation function.

In the approach of the Layzer-Irvine (LI) equation,
the kinetic energy is associated
with the absolute rms velocity (in the CMB frame) of
individual galaxies and the potential energy is an integral over
the 2-point correlation function.

{\it New developments:}
It has been realized that the pair velocity dispersion
is an unstable statistic that is dominated by galaxies in clusters.
Attempts are being made to apply the method while excluding clusters.
$\bullet$
Filtered versions of the LI approach help truncate the
(weak) divergence in estimating the potential energy and make the kinetic
energy term easier to estimate from data.

{\it Pro:}
There is no need to associate galaxies with separate groups and clusters;
it is all statistical.
$\bullet$
In the LI method the energies are dominated by contributions from the
large-scale mass distribution, and it is therefore less contaminated
by clusters and less biased by the assumption of point masses.
The two-point correlation for the LI method can be measured with
reasonable accuracy.

{\it Con:}
The relevant correlation integrals, and in particular the 3-point
correlation function that enters the CVT, are very difficult to measure.
$\bullet$
The pair velocity dispersion in the CVT is an unstable statistic,
that is dominated
by pairs of galaxies in clusters. A robust, self-consistent way to avoid
the clusters is yet to be found.
$\bullet$
The absolute velocity dispersion which enters the LI equation
is hard to measure.
$\bullet$
The CVT assumes that galaxies are point masses (or of finite size),
which overestimates the force that they exert, and biases $\omm$ low.
The overlap of extended galactic halos makes a significant difference.
$\bullet$
The methods depend on the assumption that the statistical
distribution of galaxies and mass are similar (as in the $M/L$ method).
Otherwise, they must refer to a specific model for galaxy biasing.

{\it Current Results:}
The line-of-sight pair velocity dispersion outside of clusters is in the
range $\sigma_f \simeq 300 \pm 100 \kms$ \cite{fisher94} \cite{marzke95}.
With Peebles' old estimate of the correlation functions, assuming point
masses and no biasing, he obtains from the CVT 
$\omm \simeq 0.15$ \cite{peebles97}.
$\bullet$ From
the mean pairwise velocity in IRAS 1.2Jy \cite{fisher94}:
$\omm\sim 0.25$.
$\bullet$
However, it has been demonstrated that with an extended mass distribution
in galactic halos the above CVT estimates are lower bounds,
and that the observations may be consistent with $\omm \simeq 1$
\cite{bartlett96}.

\subsection 
{Local Group Dynamics}

The mixed-boundary problem of a cosmological gravitating system
is solved for the trajectories of galaxies in the Local Group.
It is done using the least-action principle, under the assumption
that the mass is concentrated around the galaxies (\ie, strict no biasing).

{\it New Developments:}
Simulations are being used to estimate the biases in the method.

{\it Pro:}
This method is making use of accurate local measurements.

{\it Con:}
The assumption that the mass is all in point-like galaxies
during the whole evolution of the Local Group is likely to be wrong,
especially if $\omm$ is high. It causes an overestimate of the forces
and thus an underestimate of $\omm$.
This is a stronger assumption than assuming no biasing in the statistical
sense, $b=1$.
A related problem is the neglect of possible merging
that has taken place among the
initial subsystems in the Local Group.
$\bullet$
The solution to the mixed-boundary problem may be non-unique.
$\bullet$
The method treats the Local Group as an isolated system and
neglects possible tidal effects from external material.

{\it Current Results:}
Peebles \cite{peebles94} obtains $\omm \sim 0.15$, under the assumption that
galaxies strictly trace mass.
However, it has been shown using N-body simulations \cite{dunn95} that
the assumption that the mass is all in galaxies causes an underestimate
of $\omm$ by a factor $4-5$ for any true $\omm$ in the range
$0.2-1.0$.
Shaya \etal \cite{shaya95} obtain a similarly low value using a similar
analysis inside a sphere of radius $\sim 30\hmpc$.

\section 
{Large-Scale Fluctuations; Cosmic Flows}
\label{sec-flows}

These methods use the mean motions induced by the large-scale
mass distributions on scales of several megaparsecs and above.
They are based on GI in the linear regime
and approximate extensions into the mildly-nonlinear regime.
The different matter contributions to $\omm$ are relevant, 
such as ``cold" and ``hot" dark matter (see \cite{primack97} for a review), 
but the cosmological constant is not involved.
The statistical nature of the initial fluctuations has to be specified in
some cases; it is commonly assumed to be a {\it Gaussian} random field.
The initial fluctuation power spectrum is characterized by the power index $n$
on large scales, by a specific shape (\eg, as predicted by
CDM theory), and by its amplitude, \eg, via $\sigma_8$,
the \rms fluctuation in top-hat spheres of radius $8\hmpc$.
Galaxy {\it biasing} is expected to enter in a simpler manner on these larger
scales, but is still an important unknown. In the common simplified
treatments it is modeled as a linear biasing relation between the density
fluctuations of galaxies and mass, $\delta_g = b \delta$, with different
biasing parameters for different galaxy types: $b_{iras}$, $b_{opt}$, etc.
However, non-trivial properties of the biasing scheme may be important.

\subsection 
{Measured Peculiar Velocities}
\label{sec-velonly}

These methods use data that include both redshifts and redshift-independent
distances to galaxies via POTENT-like reconstructions of the underlying
velocity and mass-density fields \cite{potent}, independently of whole-sky
redshift surveys.
Combined with the assumption of Gaussian initial fluctuations,
these methods provide lower bounds on $\omm$ in various ways \cite{dekel94}, 
such as
(a) using the fact that diverging flows in voids cannot be large
when $\omm$ is low because the voids cannot be more empty than empty 
\cite{void93},
(b) using the skewness of $\divv$ as induced by $\omm$-dependent
nonlinear gravity \cite{bernardeau95}, 
and
(c) appealing to the mass-density initial probability distribution function
via a Zel'dovich ``time machine" \cite{zpom}.

{\it New developments:}
The Mark III catalog of Tully-Fisher and \dns\
peculiar velocities with more than 3000 galaxies has been completed 
\cite{mark3}, and several new samples are coming up soon.
$\bullet$
New, more accurate distance
indicators are being developed and implemented to larger samples.
Most promising are the methods based on surface-brightness fluctuations
for nearby galaxies, and supernovae type Ia for large-scale coverage
(\cite{willick97} for a review).
$\bullet$
The understanding of the systematic biases in the data and the analysis
is improving, and new methods for correcting them are being developed.

{\it Pro:} These methods allow direct dynamical measures of $\omm$ 
independent of galaxy density biasing.

{\it Con:}
The measurement of redshift-independent distances involves large errors
that grow in proportion to distance, and the analysis is therefore 
limited to the relatively local cosmological volume, out to 
$\sim 100 \hmpc$ in the best cases. The associated cosmic scatter 
is inevitably large.
$\bullet$
Some of the distance indicators may be subject to systematic environmental
effects that are hard to identify.
$\bullet$
These methods ignore possible velocity biasing.
$\bullet$
The methods assume potential flow and Gaussian initial conditions.
These are valid hypotheses for the standard scenarios of structure formations,
but they might be invalid in more esoteric models, such as those based
on topological defects as seeds for structure formation.

{\it Current Results:} From 
Mark III peculiar velocities using several methods:
$\omm > 0.3$ at more than $95\%$ confidence (reviews: \cite{dekel94}).  From 
the extracted mass power spectrum under the general CDM family of models
\cite{ps97}: $\sigma_8 \omm^{0.6} = 0.8\pm 0.2$.

\subsection 
{Redshift Distortions}

This is a statistical measure of large-scale peculiar velocities from
extended redshift surveys alone, under the assumption of global isotropy
in real space and linear biasing. The comparison of radial and angular
fluctuations yields a measure of $\beta\equiv \omm^{0.6}/b$.
There are several ways to implement this idea, using correlation
functions, power spectra, or expansion in spherical harmonics and Bessel
functions (reviews: \cite{dekel94} \cite{strauss95}).

{\it New developments:} Redshift surveys larger than before have
become available, for example PSCZ from IRAS to a flux limit of 0.6 Jy, 
and the Las Campanas Redshift Survey of optical galaxies in the south.
Even larger surveys are planned for the next few years, such as
the Two-Degree Field (2DF) and the Sloan Digital Sky Survey (SDSS)
(see \cite{strauss97}).
These catalogs will drastically reduce the cosmic scatter.

{\it Pro:} Measurements of redshifts are inexpensive; 
there is no need for independent distances. 
One can therefore use surveys of large volumes in order to beat the
measurement errors and the cosmic scatter.

{\it Con:}
In the mildly-nonlinear regime where the interpretation of
distortions is straightforward, the noise in the observations
[\eg, the $\xi(\sigma,\pi)$ diagram] is bad and difficult to quantify.
$\bullet$
With current data, the estimate suffers from large cosmic scatter.
$\bullet$
The method is subject to galaxy density biasing. At best it
measures the bias-contaminated parameter $\beta$, not $\omm$. 
The $\beta$ estimated
by this method may be systematically different from the $\beta$ estimated
by other methods \cite{dekel_lahav97}.

{\it Current Results:}
The best estimates for IRAS galaxies span a large range:
$0.5 \leq \betai \leq 1.2$ \cite{strauss97}. 
The current samples do not yet probe a sufficiently fair volume of the 
universe, and there are indications for systematic effects near the
flux limit.

\subsection 
{Velocity versus Density}

The peculiar velocity data is compared with the distribution of
galaxies in redshift space to obtain $\beta$.
The comparison can be performed
either at the density level (\eg, velocity-inferred mass density
a la POTENT versus real-space density of galaxies as extracted from
redshift surveys \cite{potiras}), 
or at the velocity level \cite{velmod} \cite{itf}, 
or simultaneously \cite{simpot}
(reviews: \cite{dekel94} \cite{strauss95}).

{\it New developments:}
The methods are being improved to better take into account the random
and systematic errors. The comparison is done in several different ways.

{\it Pro:}
Some of the comparison methods allow a direct mapping of the biasing field.
$\bullet$
Certain versions of the method are straightforward to implement.

{\it Con:}
It is hard to impose the same effective smoothing on the two data sets.
This may cause a bias in the estimate of $\beta$, and a complication due to
possible scale dependence in the biasing scheme.
$\bullet$
The estimation is contaminated by the possible complexity of the
biasing scheme.
Each method may actually refer to a somewhat different $\beta$
\cite{dekel_lahav97}.
$\bullet$
It is hard to distinguish nonlinear biasing from nonlinear gravitational
effects.

{\it Current Results:}
For IRAS galaxies, the current best estimates vary in the range
$0.5 \leq \betai \leq 1.2$, depending on the method, the volume used,
the weighting of the different data, the smoothing scale, etc.
The comparisons at the density level \cite{potiras} tend to yield higher 
estimates than the comparisons at the velocity level \cite{velmod}. 
One of the velocity comparisons indicates a possible inconsistency in the 
data at large distances \cite{itf}.
The value of $\betai$ seems to
grow with smoothing scale, from $\betai \sim 0.5-0.6$ at Gaussian smoothing
scales of $3-6\hmpc$ \cite{velmod} \cite{simpot} 
to $\betai \sim 1$ on scales of $\sim 12\hmpc$ \cite{potiras} \cite{simpot}.
The estimates for optical galaxies indicate a biasing parameter that is
typically larger by $\sim 30\%$.

\subsection 
{Cluster Abundance and Correlations}

If clusters can be modeled (\eg, using an improved version of
the Press-Schechter formalism) as ``objects" above a mass threshold in
a density fluctuation field that was initially Gaussian, then
the cluster mass function can be used to constrain $\sigma_8\omm^{0.6}$
\cite{sigma8}.
The correlation amplitude of these clusters can be compared with their
abundance to give a direct measure of $\sigma_8$.
Together, these results yield $\omm$ and $\sigma_8$ separately 
\cite{mo_white96}.

{\it Pro:}
The two parameters are determined from observational data that
are relatively easy to obtain.
$\bullet$
The method depends sensitively only on the
assumptions of Gaussian statistics and of mass-limited cluster definition.
It is insensitive to the actual power spectrum of fluctuations.

{\it Con:}
The amplitude of cluster correlations still carries a large uncertainty.
$\bullet$
The method relies on the assumption of Gaussian initial conditions.

{\it Current Results:}
$\sigma_8 \omm^{0.6} \simeq 0.5-0.6$ \cite{sigma8} 
from cluster abundances (compare to \S \ref{sec-velonly}),
but measures of the cluster autocorrelation strength are still too uncertain
to be able to give a useful second constraint \cite{mo_white96}.

\section 
{Fluctuation Growth Rate}

The amplitude of density fluctuations at redshift $z$ compared to
their amplitude today, according to linear GI theory, is approximated
to a few percent by \cite{lahav91}
\begin{equation}
\delta \simeq (1+z)^{-1}\,
2.5\,\omm\, [\omm^{4/7} -\oml +(1+\omm/2)(1+\oml/70)]^{-1} \ .
\label{eq:growth}
\end{equation}
A main discriminatory feature between low-$\omm$ and high-$\omm$
models is the effective freeze-out of the growth of fluctuations, which
occurs when the universe enters its free expansion phase; roughly
at $1+z \sim \omm^{-1}$ in an open model with $\oml =0$, 
or later at $1+z \sim \omm^{-1/3}$ in a low density flat universe with
a cosmological constant. 
Thus, structure of a given amplitude today form earlier in low-$\omm$ models
than in $\omm=1$ models, and more so in an open model than in a flat model. 
This effect can be observed in several different ways.

\subsection 
{Cluster Morphology}

Clusters are expected to be more evolved in a low-$\omm$ universe,
\ie, be more spherical and show less substructure than clusters
in an $\omm=1$ universe which must still be undergoing significant
collapse and merger activity \cite{richstone92}.

{\it New Developments:}
X-ray maps from ROSAT provide a useful measure of morphology for
many clusters.

{\it Pro:}
The method relies on a simple feature distinguishing the different
cosmologies by the rate of evolution of structure at different epochs.
Qualitative results can be obtained by visual inspection of maps.

{\it Con:}
The method requires high-resolution simulations with gas dynamics
and many clusters in order to beat the cosmic scatter.
$\bullet$
Substructure is hard to quantify.
$\bullet$
The substructure depends also on the fluctuation power spectrum,
on $\omb$, and on the various other dark matter species.

{\it Current Results:}
Dissipative simulations \cite{mohr95} agree
with dissipationless simulations \cite{jing95} 
that a significant effect is expected. However,
these papers disagree about its strength and about
the conclusions to be drawn from comparison with the clusters observed
by ROSAT. The current situation is somewhat confused.





\subsection 
{The Epoch of Galaxy Formation}

Can the observed number count $N(z)$ at high $z$
(of a few) be compatible with $\omm=1$?
This test can involve various objects such as quasars, early galaxies,
damped Lyman alpha systems. etc.

{\it New Developments:}
There are HST and Keck observations of central regions of galaxies in
an early stage of their formation at $z \sim 3-3.5$ (some claim that
galaxies are observed all the way to $z\sim 6$ based on ``photometric
redshifts"). Improved spectrographs are being developed, as well
as methods for estimating ``photometric redshifts".

{\it Pro:}
New data are accumulating rapidly from HST and 8-10 meter telescopes.

{\it Con:}
The method is contaminated by unknown evolutionary issues.
$\bullet$
The model predictions depend on the power spectrum of fluctuations.
$\bullet$
Dissipationless simulations predict the number density of halos $N(M,z)$,
but not necessarily of luminous objects.

{\it Current Results:}
The number count of quasars may be consistent with $\omm=1$
provided that there was efficient cooling and angular-momentum
transfer \cite{katz94}.
$\bullet$
The abundance of bright galaxies seen at $z\sim 3.5$
\cite{steidel96} may favor low $\omm$ \cite{mo_hdf96}.
On the other hand, the relatively small number of galaxies identified
in the Hubble Deep Field
with colors consistent with very high redshifts ($z=4-6$)
\cite{lanzetta96} may indicate that low $\omm$ values can be excluded.

\subsection 
{Present Structure versus Fluctuations in the CMB}

This method is
using independent constraints on the power spectrum of today's
fluctuations on scales $\sim 100\hmpc$, and the power spectrum of
fluctuations at $z\sim 10^3$ on scales $100-1000\hmpc$, assuming
gravitational growth.

{\it New Developments:}
Peculiar velocity data enable constraints on $\omm$ independent of biasing.
Future extended redshift surveys (2DF, SDSS) will provide
constraints on $>100\hmpc$ scales.
COBE data put limits on scales $\sim 1000\hmpc$.
Accumulating ground-based and balloon data of CMB fluctuations
with sub-degree resolution start providing constraints on
scales $\sim 100\hmpc$.

{\it Pro:}
With peculiar velocity data this method compares measures of dynamical
fluctuation fields independent of galaxy biasing.

{\it Con:}
As long as the scales explored by today's structure and the CMB
fluctuations do not overlap, the constraint on $\omm$ depends on $n$.
$\bullet$
A hot dark matter component would alter the result via a different
fluctuation growth rate, and confuse the constraints.
$\bullet$
If today's power spectrum is extracted from a galaxy redshift survey
then the method depends on galaxy biasing on large scales.

{\it Current Results:}
Using COBE and Mark III velocities,
and considering the family of Inflation-motivated
CDM models with a possible cosmological
constant such that $\omm+\oml=1$ and a possible tilt in the
initial power spectrum, a likelihood analysis yields:
$\omm \h65 n^2 = 0.7 \pm 0.1$ \cite{ps97}.
The best fit for CDM is thus obtained with a slight deviation from the
``standard" CDM model, of
either $n\sim 0.8-0.9$, $\omm \sim 0.7$, or $\omn \sim 0.2$.
The indicated height of the first acoustic peak of the CMB
allows only a slight tilt, $n \sim 0.9$, and a high baryon content,
$\omb \sim 0.1$ \cite{zaroubi_cmb97}.







\section
{Conclusion}

Table 1 summarizes the current estimates of $\omm$ from the various
methods of measurement. In general, the methods based on virialized objects
favor low values, while the global measures (with the exception of 
the age argument) and the analysis of large-scale structure and flows 
tend towards higher values.
The estimates that indicate very low values ($\sim 0.1-0.2$)
have plausible loopholes. On the other hand, none of the estimates that favor
high $\omm$ values actually requires that $\omm$ is as large as unity.
It thus seems that a tentative consensus can be reached at a (not very
elegant) value of about $\omm \simeq 0.5$.
This was reflected in the debate on formation scenarios by the fact that
the competing CDM models were the Einstein deSitter $\omm=1$
``standard'' CDM (with $n\sim 0.9$) or Hot+CDM, versus the Open CDM
and Flat ($\Lambda$) CDM models with $\omm \simeq 0.4$.
Not too long ago these low-$\omm$ models used to be associated
with $\omm \sim 0.1-0.2$.  It seems that progress is being made towards
convergence.
The activity on many fronts of this field promises
that we will know more in the near future. 
We hope that the above discussion will be of help in the effort to
reconcile the various estimates with a unique value of $\omm$.


 \bigskip\bigskip

\def \srule {
        \vskip 0.4\baselineskip
        \hrule height.7pt
        \vskip 0.3\baselineskip }
\def \drule {
        \vskip 0.4\baselineskip
        \hrule height.7pt
        \vskip2pt
        \hrule height.7pt
        \vskip 0.3\baselineskip }
 
\cl{\bf Table 1: Estimates of \pmb{$\omm$}}
\smallskip
 
\vbox {
\drule
\halign to \hsize {
#\quad\hfil&#\quad\hfil&#\hfil\cr
 
&&\cr
Global Measures
&Inflation, Occam
&$\omm+\oml=1$ \ ($\omm=1$, $\oml=0$)          \cr
&Lum. distance SNIa
&$-0.3<\omm-\oml<2.5$ (90\%) \
\cr
&&Flat $\omm > 0.49$ (95\%)     \cr
&Lens Counts
&Flat $\omm > 0.34$ (95\%)                        \cr
&CMB Peak
&$\omm + \oml < 1.5$ (95\%) \cr
&
&$\omm + \oml > 0.3$ \ (likely $\sim 0.7$)         \cr
&$\ho\to$
&$\omm -0.7\oml < 1.3$ \ (likely $\leq 0$)   \cr
 
&&\cr
Virialized Objects
%
&$(M/L){\cal L}$
&$\omm \sim 0.25$  ($0.1-1.0$)                        \cr
&Baryon fraction
&$\omm \h65^{1/2} \sim 0.3-0.5$ (low$-$high $\omb$) \cr
&Cosmic Virial Th.
&Point mass $\omm \!\sim\! 0.2$ (halos $\!\rightarrow\!1$)      \cr
&Local Group
&Point mass $\omm \!\sim\! 0.15$ (halos $\!\rightarrow\!0.7$)  \cr
 
&&\cr
Large-Scale; Flows
&Peculiar velocities
&$\omm > 0.3$ (95\%) \cr
&
&$\omm^{0.6}\sigma_8^a =0.8\pm 0.2$ \ ($\betai^b \simeq 1.05^c$)\cr
&Redshift Distortions
&$\betai \sim 0.5-1.2$  \cr
&Velocity vs Density
&$\betai \!\sim 0.5-1.2$ (scale dependent) \cr
&&$\betao \!\sim 0.4-0.95$   \cr
&Cluster Abundance
&$\omm^{0.6} \sigma_8 \simeq 0.5-0.6$ \ ($\betai \simeq 0.7-0.8^c$)\cr
 
&&\cr
Fluct. Growth
%
&Cluster Morphology
&$\omm > 0.2$ (?)         \cr
%
%
&Galaxy Formation
&(?) \cr
&$P_k(\rho)$ vs $C_l$
&CDM $n=1$ $b=1$: $\omm \h65 \sim 0.3$  \cr
&$P_k(v)$ vs $C_l$
&CDM flat: $\omm \h65 n^2 \simeq 0.7 \pm 0.1$  \cr
&& \cr
\noalign {\srule}
&&\cr
}
}
\no$^a$ $\sigma_8$ is the {\it rms} mass density fluctuation in a top-hat
    sphere of radius $8\hmpc$. \bk
\no$^b$ $\beta\equiv\Omega^{0.6}/b$,\ \ ~$\bi$ for IRAS galaxies,
                         ~ $\bo$ for optical galaxies. \bk
\no$^c$ $\bo/\bi \simeq 1.3$,\ \ ~$\bo\simeq 1/\sigma_8$.


\vfill\eject


\end{document}